\documentclass[12pt]{iopart}
\usepackage{graphicx}
\usepackage{mathrsfs}
\usepackage{multirow}
\usepackage{epsfig}
\usepackage{color}

\usepackage{iopams,epsfig}
\newcommand{\bea}{\begin{eqnarray}}
\newcommand{\eea}{\end{eqnarray}}
\newcommand{\ba}{\begin{array}}
\newcommand{\ea}{\end{array}}

\newcommand{\be}{\begin{equation}}

\newcommand{\ee}{\end{equation}}
\newcommand{\bt}{\begin{teo}}
\newcommand{\et}{\end{teo}}

\newcommand{\la}{\lambda}

\newcommand{\al}{\alpha}

\newcommand{\erf}{{\rm erf}}
\newcommand{\erfc}{{\rm erfc}}

\begin{document}

\title[Dynamical localization in chaotic systems]
{Dynamical localization of chaotic eigenstates in the mixed-type systems: 
spectral statistics in a billiard system after separation of regular
and chaotic eigenstates}

\author{Benjamin Batisti\'c and Marko Robnik}

\address{CAMTP - Center for Applied Mathematics and Theoretical Physics,
University of Maribor, Krekova 2, SI-2000 Maribor, Slovenia, European Union}


\eads{Benjamin.Batistic@gmail.com, Robnik@uni-mb.si}

\begin{abstract}
We study the quantum mechanics of a billiard (Robnik 1983) in
the regime of mixed-type classical  phase space (the shape parameter $\la=0.15$)
at very high-lying eigenstates, starting at about 1.000.000th
eigenstate and including the consecutive 587654 eigenstates. 
By calculating  the normalized Poincar\'e Husimi functions of the 
eigenstates and comparing them with the classical phase space 
structure, we introduce the overlap criterion which enables
us to separate with great accuracy and reliability the regular and chaotic
eigenstates, and the corresponding energies. The chaotic eigenstates
appear all to be dynamically localized, meaning that they
do not occupy uniformly the entire available chaotic classical
phase space component, but are localized on a proper subset. We find 
with unprecedented precision and statistical significance 
that the level spacing distribution of the regular levels obeys
the Poisson statistics, and the chaotic ones obey the Brody
statistics, as anticipated in a recent paper by Batisti\'c 
and Robnik (2010), where the entire spectrum was found to
obey the BRB statistics. There are no effects of dynamical 
tunneling in this regime, due to the high energies, as they
decay exponentially with the inverse effective Planck constant
which is proportional to the square root of the energy.
\end{abstract}

\pacs{01.55.+b,02.50.Cw,02.60.Cb,05.45.Pq, 05.45.Mt, 47.52.+j}

\submitto{\JPA} 


\section{Introduction} \label{Intro}

In quantum chaos \cite{Stoe,Haake,Rob1998} of general (generic) 
time-independent 
(autonomous) Hamilton systems in the strict semiclassical 
limit we can conceptually separate regular and chaotic 
eigenstates. This picture goes back to the work by Percival in 1973
\cite{Per1973}
and Berry and Robnik in 1984 \cite{BR1984}. One of the main 
results in quantum chaos is the fact that
in classically fully chaotic  (ergodic,
autonomous Hamilton) systems with the purely discrete spectrum
the fluctuations of the energy spectrum around its mean behaviour
obey the statistical laws described by the Gaussian
Random Matrix Theory (RMT) \cite{Mehta,GMW}, provided that we are in the
sufficiently deep semiclassical limit. The latter condition
means that all relevant classical
transport times, like the typical ergodic time, or diffusion time,
 are smaller than the so-called Heisenberg time,
or break time, given by $t_H=2\pi\hbar/\Delta E$, where
$h=2\pi\hbar$ is the Planck constant and $\Delta E$ is the mean
energy level spacing, such that the mean energy level density
is $\rho(E) =1/\Delta E$. This statement is known as the Bohigas -
Giannoni - Schmit (BGS) conjecture and goes back to their pioneering
paper in 1984 \cite{BGS}, although some preliminary ideas
were published in \cite{Cas}.  Since $\Delta E \propto \hbar^d$,
where $d$ is the number of degrees of freedom (= the dimension of
the configuration space), we see that for sufficiently small
$\hbar$ the stated condition will always be satisfied.
Alternatively, fixing the $\hbar$, we can go to high energies
such that the classical transport times become smaller than
$t_H$. The role of the antiunitary symmetries that classify
the statistics in terms of GOE, GUE or GSE (ensembles of RMT) has been
explained in \cite{RB1986}, see also \cite{Rob1986}
and \cite{Stoe,Haake,Rob1998,Mehta}. The theoretical foundation
for the BGS conjecture has been initiated first
by Berry \cite{Berry1985},  using the Gutzwiller periodic 
orbit theory (trace formula) \cite{Gutz} (for an excellent exposition
see \cite{Stoe})  and later further developed
by Richter and Sieber \cite{Sieber}, arriving finally in
the almost-final proof proposed by the group of F. Haake
\cite{Mueller1,Mueller2,Mueller3,Mueller4}.

On the other hand, if the system is classically
integrable, Poisson statistics applies, as is well known
and goes back to the work by Berry and Tabor in 1977 (see
\cite{Stoe,Haake,Rob1998} and the references therein,
and for the recent advances \cite{RobVeb}).

In the mixed-type regime, where classical regular regions
coexist in the classical phase space with the chaotic regions,
being a typical KAM-scenario which is the generic situation,
the so-called Principle of Uniform Semiclassical Condensation
(of the Wigner functions of the eigenstates; PUSC) applies, based
on the ideas by Berry \cite{Berry1977}, and further extended by
Robnik \cite{Rob1998}. If the  stated semiclassical condition
is satisfied, the chaotic eigenstates are uniformly extended,
and  consequently the Berry-Robnik statistics \cite{BR1984,
ProRob1999}  is observed - see also \cite{Rob1998}.
If the semiclassical condition stated above requiring that
$t_H$ is larger than all classical transport times is not
satisfied, the chaotic eigenstates will not be extended but
localized
and the Berry-Robnik statistics must be generalized as
explained in \cite{ProRob1994a,ProRob1994b,Rob1998,BatRob2010,VebRobLiu1999}
and in this paper.

The relevant papers dealing with the mixed-type regime
after the work \cite{BR1984} are
\cite{ProRob1999} - \cite{GroRob2} and the most recent
advance was published in \cite{BatRob2010}.
If the couplings
between the regular eigenstates and chaotic eigenstates
become important, due to the dynamical tunneling, we
can use the ensembles of random matrices that capture these
effects \cite{VSRKHG,BatRob2010}.  As the tunneling strengths typically decrease
exponentially with the inverse effective Planck constant,
they rapidly disappear with increasing energy, or by
decreasing the value of the Planck constant. In this
work we shall deal only with high-lying eigenstates,
and therefore we can neglect the effects of tunneling.

However, quite generally, if the semiclassical condition is not satisfied,
such that $t_H$ is no longer larger than the relevant classical
transport time, like e.g. the diffusion time in fully chaotic but
slowly ergodic systems, we find the so-called {\bf dynamical
localization}, or {\bf Chirikov localization}. 
Dynamical localization was discovered in time dependent
systems \cite{CCFI79}.  It was intensely studied since then
in particular by Chirikov, Casati, Izrailev, Shepelyanski
and Guarneri, in the case of the kicked rotator as reviewed in
\cite{Izr1990}. See also the references  \cite{Izr1988}-\cite{Izr1989},
and the most recent work \cite{MR2012}. 
For a general overview of the time dependent Floquet systems
see also \cite{Stoe, Haake}. It has been 
observed that in parallel with the localization of the eigenstates
one observes the fractional power law level repulsion (of the quasienergies)
even in fully chaotic regime (of the finite dimensional 
kicked rotator), and it is believed
that this picture applies also to time independent (autonomous)
Hamilton systems and their eigenstates \cite{MR2012}.
(See the excellent review of localization in time independent
billiards by Prosen in \cite{Pro2000}.)
Indeed, this has been analyzed with
unprecedented precision and statistical significance recently by
Batisti\'c and Robnik \cite{BatRob2010} in case of mixed-type 
systems, and the present work is being extended in 
the analysis of separated regular and chaotic eigenstates.
An early attempt of separation of eigenstates in the
billiard system has been published in \cite{LiRob1995},
using a different approach at much lower energies
and with much smaller statistical significance.
In \cite{BB2007} mushroom billiards were studied at much
lower energies and with much smaller statistical significance,
where the aspects of tunneling were investigated in the first
place, but not the dynamical localization, although a
clear deviation from the GOE statistics was found in
the chaotic eigenstates. 

In this paper we introduce a criterion for classifying
eigenstates as regular and chaotic,
and moreover, we show that the regular levels obey the
Poisson statistics, whilst
the chaotic dynamically localized
eigenenergies obey exceedingly
well the Brody distribution \cite{Bro1973}, with the
Brody parameter values $\beta$ within the interval
$[0,1]$, where $\beta=0$ yields the Poisson distribution
in case of the strongest localization, and $\beta=1$ gives
the Wigner surmise (2D GOE, as an excellent approximation of
the infinite dimensional GOE), which describes the 
extended chaotic eigenstates. It turns out that
the Brody distribution introduced in \cite{Bro1973},
see also \cite{Bro1981},
fits the empirical data much better
than e.g. the distribution function proposed by F. Izrailev
(see \cite{Izr1988,Izr1990} and the references therein).

It is well known that Brody distribution so far has no
theoretical foundation, but our empirical results
show that we have to consider it seriously  in dynamically
localized chaotic eigenstates, thereby
being motivated for seeking its physical foundation,
and an analogous result was obtained in the recent work
of Manos and Robnik (2013) \cite{MR2012} in the analysis
of the quantum kicked rotator, where the object of study
are the eigenstates of the Floquet operator and the
statistical properties of the spectrum of eigenphases
(quasienergies) in classically fully chaotic regime.

In the Hamilton systems with classically mixed-type dynamics, which is the
generic case,  we have classically regular quasi-periodic motion
on $d$-dim invariant tori ($d$ is the number of freedoms) for
some initial conditions (with the fractional Liouville volume $\rho_1$)
and chaotic motion for the complementary initial conditions
(with the fractional Liouville volume $\rho_2=1 - \rho_1$).
The chaotic set might be further decomposed into several
chaotic regions (invariant components) in case $d=2$, whilst
for $d>2$ it is strictly speaking always just one chaotic set
due to the Arnold diffusion on the Arnold web, which pervades the
entire phase space. In sufficiently deep semiclassical
limit the Berry-Robnik picture \cite{BR1984} is established, based on the
statistically independent superposition of the regular
(Poissonian) and chaotic level sequences, based on PUSC, as
explained above.

In this paper we shall
consider only the case of just one chaotic component, although
the results can be easily generalized for more than one chaotic
component. This picture gives an excellent approximation
for the statistics of spectral fluctuations of the mixed-type
systems, if the largest chaotic component is much larger than
the next largest one, which typically indeed is the case e.g. in
2D billiards. 

The energy spectrum of the mixed-type system with one regular and
one chaotic component can
be described in the Berry-Robnik (BR) regime of sufficiently small
effective Planck constant $\hbar_{eff}$ by the following formula 
for the gap probability  $E(S)$, 

\be \label{BRE} 
E(S) =  E_r(\rho_1 S) E_c(\rho_2 S)
\ee
and the level spacing distribution $P(S)$ (see e.g. \cite{Rob1998})
is of course always given as the second derivative of the gap
probability, namely $P(S) = d^2 E(S)/dS^2$, so that we have

\be \label{BRP} 
P(S) =  \frac{d^2}{dS^2} E_r(\rho_1 S) E_c(\rho_2 S) =
\frac{d^2E_r}{dS^2} E_c + 2\frac{dE_r}{dS} \frac{dE_c}{dS} + 
E_r \frac{d^2E_c}{dS^2}.
\ee
This factorization formula (\ref{BRE}) is a direct consequence
of the statistical independence, justified by PUSC. Here by
$E_r(S) = \exp (-S)$ we denote the gap probability for the Poissonian
sequence with the mean level density one. By $E_c(S)$ we denote the
gap probability for the chaotic level sequence with the mean
level density (and spacing) one. Note that the classical parameter
$\rho_1$ and its complement $\rho_2=1-\rho_1$ enter the expression
as weights in the arguments of the gap probabilities.

Using the Bohigas-Giannoni-Schmit conjecture we conclude that in
the sufficiently deep semiclassical limit $E_c(S)$ is given by
the RMT, and can be well approximated by the Wigner surmise

\be \label{PCW} 
P_W(S) = \frac{\pi S}{2} \exp \left(-\frac{\pi S^2}{4} \right),\;\; 
F_W(S) = 1 - W_W(S) = \exp \left( -\frac{\pi S^2}{4} \right), 
\ee
such that $E_c(S)$ is equal to

\be  \label{ECW}
E_W(S) =  1 - \erf \left( \frac{\sqrt{\pi}S}{2} \right) = 
\erfc \left( \frac{\sqrt{\pi}S}{2} \right),
\ee
where $\erf (x) = \frac{2}{\sqrt{\pi}}\int_0^x e^{-u^2} du$ is the error
integral and $\erfc (x)$ its complement, i.e. $\erfc (x) = 1 - \erf (x)$.
In the equation (\ref{PCW}) $W_W(S)$ denotes the cumulative Wigner level
spacing distribution $W_W(S) = \int_0^S P_W(x)\;dx$ and $F_W$ its
complement. The explicit Berry-Robnik level spacing
distribution (in the special case
of one regular and one chaotic component) follows immediately,

\be \label{BRexplicit}
P_{ BR}(S) = e^{-\rho_1 S} \left\{ e^{- \frac{\pi \rho_2^2
S^2}{4}} \left(2 \rho_1 \rho_2 + \frac{\pi \rho_2^3 S}{2} \right) + \rho_1^2 {\rm erfc}
\left(\frac{\sqrt{\pi} \rho_2 S}{2}\right) \right\}. 
\ee
The correctness of this distribution function in the BR
regime (sufficiently small $\hbar_{eff}$) is by now very well established
in highly accurate numerical calculations for all
$E(k,L)$ probabilities, not only the gap probability \cite{ProRob1999}.

In the present work the above basic BR formula (\ref{BRE}) is
generalized as in \cite{BatRob2010} to capture the dynamical localization
effects, responsible for the deviation from the BR regime.

At not sufficiently small $\hbar_{eff}$ (e.g. in billiards this means 
at low energies) the chaotic eigenstates (their Wigner functions in the
phase space) are not uniformly extended over the entire classically allowed
chaotic component, but are dynamically localized.
Thus we see the transition
from GOE in case of extended chaotic states to the
Poissonian statistics in case of strong localization.
The level spacing distribution in such a transition regime of
localized chaotic eigenstates can
be described by the  Brody distribution with the
only one family parameter $\beta$,

\be \label{BrodyP}
P_B(S) = C_1 S^{\beta} \exp \left( - C_2 S^{\beta +1} \right), \;\;\; 
W_B(S) = 1-  \exp \left( - C_2 S^{\beta +1} \right),
\ee
where the two parameters $C_1$ and $C_2$ are determined by the two
normalizations $<1> = <S> = 1$, and are given by

\be \label{Brodyab}
C_1 = (\beta +1 ) C_2, \;\;\; C_2 = \left( \Gamma \left( \frac{\beta +2}{\beta +1}
 \right) \right)^{\beta +1}
\ee
with  $\Gamma (x)$ being the Gamma function. As mentioned before,
if we have extended
chaotic states $\beta=1$ and RMT  (\ref{PCW}) applies, whilst in the strongly 
localized regime $\beta=0$ and we have Poissonian statistics.
Again, by $W_B(S)$ we denote the cumulative Brody level spacing
distribution, $W_B(S) = \int_0^S P_B(x)\;dx$.
The corresponding gap probability is 

\be \label{BrodyE}
E_B(S) = \frac{1}{ (\beta +1) \Gamma \left(\frac{\beta +2}{\beta +1}\right) }
  Q \left( \frac{1}{\beta +1}, \left( \Gamma\left(\frac{\beta +2}{\beta +1}
 \right) S \right)^{\beta +1} \right)
\ee
where  $Q(\al, x)$ is the incomplete Gamma function

\be \label{IGamma}
Q(\al, x) = \int_x^{\infty} t^{\al-1} e^{-t} dt.
\ee
By choosing $E_c(S)$ in equation (\ref{BRE}) as given in (\ref{BrodyE})
we are able to describe the localization effects on the chaotic component.
Such approach has been already proposed in the paper by Prosen and Robnik
in 1994 \cite{ProRob1994a, ProRob1994b} and the resulting level spacing distribution, 
emerging from this assumption, was called Berry-Robnik-Brody (BRB).
It has two parameters, the classical parameter $\rho_1$ and the quantum
parameter $\beta$.
It will turn out that this description is indeed excellent,
and has been verified to high accuracy by Batisti\'c and Robnik \cite{BatRob2010}.

The applicability of the Brody distribution in this context is 
theoretically still not
well understood, but we shall see that the theory describes
very well the empirical data from the highly accurate energy spectra
of billiards at energies around and below the deep semiclassical
(Berry-Robnik) regime. Therefore,
the resulting theory is of semiempirical nature, but seems to be
quasi-universal in the sense that below the Berry-Robnik regime we
indeed find spectral fluctuations, in particular the level spacing distribution,
which are well described by our theory on the very finest 
scale of level spacings. 
We have also tried to use other one-parametric level spacing 
distributions instead of Brody, in particular those proposed and 
studied in Izrailev's papers \cite{Izr1988,Izr1989,Izr1990} 
(see also \cite{CIM1991}), 
but  must definitely conclude that the Brody distribution is quite
special, it gives by far the best agreement between the theory and real 
spectra. Izrailev's intermediate level spacing distribution intended
to capture the localization effects manifested in the quantal spectra
is given by

\be \label{IzrailevDistrib}
      P_I(S) = A\left (\frac{1}{2}\pi S \right )^{\beta} \exp \left [-\frac{1}{16}\beta \pi^2 S^2 -\left (B-\frac{1}{4}\pi \beta \right )S \right],
\ee
where the constants $A$ and $B$ are determined by the normalizations $<1>=<S>=1$.

As we shall show the dynamical localization effects can persist up
to very high-lying eigenstates, even up to one million, 
whilst - as mentioned before - 
the tunneling effects occur usually only at very low-lying eigenstates,
due to the exponential dependence on the reciprocal effective
Planck constant, $\propto \exp(-const./\hbar_{eff})$, and thus
can be neglected in our case.
   
The paper is structured as follows: In section \ref{Model} the billiard
system is defined as introduced by Robnik 
\cite{Rob1983,Rob1984} with shape parameter $\lambda=0.15$, 
and we describe the Poincar\'e Husimi functions, in section \ref{Separ} 
we introduce the method of classifying and separating regular and chaotic 
eigenstates in terms of normalized Poincar\'e Husimi functions (which are 
Gaussian-smoothed Wigner functions), in section \ref{Results} 
we present the results,
and in section \ref{Conclusion} we conclude and discuss the results.

\section{Introducing the model system and the definition of the problem} \label{Model}

As an interesting and frequently studied model system we have chosen
the billiard introduced by Robnik in references \cite{Rob1983}-\cite{Rob1984},
whose boundary is defined by the quadratic conformal map of the
unit circle $|z|=1$ of the $z$-complex plane onto the $w$-complex plane
(which is the physical plane) as follows

\be \label{defbil}
w= z + \la z^2.
\ee
The choice of this quadratic map has two reasons: (i) it allows
for an elegant method \cite{Rob1984} to solve the Helmholtz equation in the
$w$-plane by transforming back to the $z$-plane, and (ii) it is the simplest
one with nontrivial classical dynamics  \cite{Rob1983}.
The shape parameter $\la$ goes from 0 (circle; integrability) to 1/2 
(cardioid billiard; ergodicity: full chaos \cite{Mar1993}). For $0\le \la \le 1/4$
the billiard boundary is convex and thus we observe the existence of
Lazutkin's caustics \cite{Laz1981,Laz1991} near the boundary in the configuration space, 
which are the projections of the KAM invariant tori (in the phase space).
When increasing the value of $\la$ from 0 we have at $\la=1/4$ for the first 
time a point of zero curvature located at $z=-1$, and thus $w=-1+\la$. 
By a theorem due to J. Mather this is a sufficient condition for the
destruction of Lazutkin's caustics and of the underlying invariant tori
near the boundary, which in turn is a necessary condition for the
ergodicity of the classical billiard dynamics. However, at $\la \ge 1/4$
the dynamics is not yet ergodic, fully chaotic, as there are still
some tiny KAM islands of stability \cite{Hayli1987}, not so easy to detect
numerically.  Nevertheless, for $\la=1/2$, the ergodicity was 
proven rigorously by Markarian \cite{Mar1993}. 

We are interested in the mixed-type case, within the interval $0 < \la <1/4$,
where the chaotic regions coexist with the regular regions in the phase space.
In particular, we have chosen $\la =0.15$, which is one of the most
frequently studied cases in \cite{ProRob1993a,ProRob1993b,ProRob1994a,
ProRob1994b,VebRobLiu1999,Dob1996,LiRob1994,LiRob1995}.
The fractional phase space volume $\rho_1$ of the classically regular
part of the phase space (not to be confused with the area on 
the Poincar\'e surface of section!) is
equal to $0.175$, as has been carefully studied in \cite{Dob1996,BatRob2010}.
In a previous work \cite{ProRob1994a,ProRob1994b}  $\rho_1$ was
estimated numerically as $\rho_1=0.36$, which is due to the technical
difficulties in distinguishing the regular regions and slowly diffusing
chaotic regions due to the sticky objects in the classical phase space.
These difficulties were overcome in \cite{Dob1996} and \cite{BatRob2010},
using the new methods based on the ideas and approach in \cite{RobDob1997}.  
The estimate $\rho_1=0.175$ is now believed to be accurate within at
least one percent relative error.

The classical
mechanics of 2D billiards is studied in the Poincar\'e-Birkhoff 
coordinates $(s,p)$, where $s$ is the  arclength
parameter going from 0 to ${\cal L}$, the perimeter of the
billiard domain, in our case counted anticlockwise from the point
$w=1+\la$, and $p$ is simply the sine of the reflection angle, 
which goes from -1 to +1. The bounce map is defined by the free
motion between the collision points on the boundary, obeying the
specular reflection law upon each collision. Thus, 
the complete information on the classical dynamics is contained
in the structure of the bounce map on the cylinder $(s,p)$. 
When analyzing the quantum mechanics
of this system, we would like to find an analogous two dimensional 
space which
also contains the complete information about the quantum mechanics,
namely about the eigenfunctions.
This is the space of the so-called Poincar\'e Husimi functions
(see \cite{BFS2004} and the references therein) that we introduce
below.

The quantum mechanics of the billiard system comprises the study of the
solution of the Helmholtz equation for the billiard domain ${\cal B}$,

\be \label{Helm}
\Delta \psi + k^2 \psi =0,
\ee
with the Dirichlet boundary condition $\psi=0$ on the boundary $\partial{\cal B}$.
We have used a number of methods, like in \cite{BatRob2010}, to calculate the
eigenenergies $E_j=k_j^2$, where $j$  is the counting index $j=1,2,3,\dots$, 
and the associated eigenfunctions are denoted  by  $\psi_j ({\bf r})$. 

Introducing the important quantity $u(s)$, the normal derivative
of the eigenfunction  $\psi$ on the boundary, from here onwards
called {\em boundary function},

\be \label{normalder}
u(s) = {\bf n}\cdot \nabla_{{\bf r}} \psi \left({\bf r}(s)\right),
\ee
where ${\bf n}$ is the unit outward vector  normal to the boundary
at position $s$, we can show \cite{BerWil1984} that 
the eigenvalue problem (\ref{Helm}) is equivalent to the following
integral equation

\be  \label{HelmInt}
u(s) = -2 \oint dt\; u(t)\; {\bf n}\cdot\nabla_{{\bf r}} G({\bf r},{\bf r}(t)).
\ee
Here ${\bf r}(t)$ is the position vector at the point $s=t$ on
the boundary, whilst ${\bf r}$ is the position vector inside the
billiard ${\cal B}$.  $G({\bf r},{\bf r}(t))$ is the free particle Green
function, namely 

\be \label{Green}
G({\bf r},{\bf r'}) = -\frac{i}{4} H_0^{(1)}(k|{\bf r}-{\bf r'}|),
\ee
where $H_0^{(1)} (x)$ is the zero order Hankel function of the first kind.
It is important to know that knowing $u_j(s)$ for a certain eigenfunction
$j$, we can immediately calculate the wave function $\psi_j({\bf r})$
within the interior of the billiard domain by 

\be \label{Intfromu}
\psi_j({\bf r})  = - \oint dt\; u_j(t)\; G\left({\bf r},{\bf r}(t)\right).
\ee
Thus, in certain analogy to the classical mechanics, the
quantum mechanics is completely described by the boundary
functions $u_j(s)$. Finally, there is the important identity \cite{BerWil1984}

\be \label{Intu2}
\frac{1}{2} \oint dt\; {\bf n}(t)\cdot{\bf r}(t)\; u_j(t)^2 = k_j^2.
\ee
The quantum analogy of the classical phase space is the 
space of Wigner functions \cite{Wig1932} or of other phase space 
representations of quantum states in general. 
The Wigner functions
are real but not positive definite functions, and exhibit lots
of oscillations around the zero level also in the regime where the 
quantum probability density is low and thus such
structures often obscure the main physical aspects of the phenomena.
Nevertheless, they do uniformly condense on the classical invariant 
objects, according to the mentioned PUSC \cite{Rob1998} in the introduction.
There are different ways of defining positive definite phase space
functions, but Husimi functions \cite{Hus1940} are perhaps the best way to
do it. They are in fact Gaussian smoothed Wigner functions.
In general, we define them by the projection of the wave function
onto a coherent state. One of the possible formulations can be found
in \cite{BFS2004}, whose definitions and the  notation we shall
use in what follows. The idea and the approach (in slightly different
form)  goes back to the works \cite{CPC1993},
\cite{TV1995} and \cite{SVS1997}. The most important idea is to define
the one-dimensional coherent states onto which we project the
boundary functions $u_j(s)$. For this reason, and due to the analogy
with the classical dynamics and its Poincar\'e surface of section
on the {\em cylinder} $(s,p)$, the underlying Husimi functions are
called {\em Poincar\'e Husimi functions} \cite{BFS2004}.

The key idea is to introduce one-dimensional coherent state as a function
of the coordinate $s$ on the boundary $\partial {\cal B}$,
localized at $(q,p)\in [0,{\cal L}]\times \mathbb{R}$, 
which is properly {\em periodized}, as introduced
by Tualle and Voros \cite{TV1995}, but here following the notation
from \cite{BFS2004} we define

\be \label{cohsta}
c_{(q,p),k} (s) = \sum_{m\in \mathbb{Z}} 
\exp \{ i\,k\,p\,(s-q+m{\cal L})\} 
\exp \left(-\frac{k}{2}(s-q+m{\cal L})^2\right).
\ee
The periodicity in $s$ with the period ${\cal L}$ is now obvious.
Here we have dropped all normalization factors, because in the
end we shall normalize the Poincar\'e Husimi functions anyway. 
Then, using this, the Poincar\'e Husimi function associated with
the $j$-th eigenstate represented by the boundary function $u_j(s)$
with the eigenvalue $k=k_j$, is

\be \label{Husfun}
H_j(q,p) = \left| \int_{\partial {\cal B}} c_{(q,p),k_j} (s)\;
u_j(s)\; ds \right|^2 ,
\ee
which is positive definite by construction. In the semiclassical limit
$j \rightarrow \infty$, and $k_j\rightarrow \infty$, we shall observe
that the Poincar\'e Husimi function is concentrated on the classical 
invariant regions, which can be an invariant torus,
a chaotic component, or the entire Poincar\'e surface of section $(s,p)$
if the motion is ergodic. This is a consequence of PUSC, bearing in
mind that the Husimi function is just a Gaussian smoothed  Wigner
function, where in the semiclassical limit the width of the smoothing Gaussian
becomes less and less important. Therefore, we expect that in the 
semiclassical limit the Poincar\'e Husimi functions will directly
correspond to either the classical regular regions or to the 
classical chaotic regions, with the exceptions having measure zero.
We can then use the Poincar\'e Husimi functions to classify and thus
also to separate the regular and chaotic eigenstates, and thereby 
also separate the regular and the chaotic spectral subsequences of the
energy eigenvalues $E_j=k_j^2$. This is what we do in the next section
\ref{Separ}.

\section{Separating the regular and chaotic eigenstates and subspectra} \label{Separ}

We consider the billiard (\ref{defbil}) with $\la=0.15$. As mentioned,
the value of the classical parameter is $\rho_1=0.175$.
The quantum eigenstates were calculated using the method of Vergini
and Saraceno \cite{VS1995} with great accuracy, for all eigenstates
(587654) within the interval $k\in[2000,2500]$. The number of eigenstates
below $k=2000$ is estimated by the Weyl rule as about 1.000.000.
In Appendix A we show that the semiclassical condition of
sufficiently large ratio of the classical transport time $t_T$ and
the Heisenberg time $t_H$ is well satisfied, satisfying the
inequality $k \ll N_T/2$, where $N_T$ is the classical transport time
in units of the number of collisions, and is equal to $N_T\approx 10^5$.
Then, when calculating
the Poincar\'e Husimi functions, the momentum $p$ is rescaled 
by the eigenvalue $k_j$, such that $p=1$ corresponds to the
original $p=k_j$. For each eigenstate
the Poincar\'e Husimi function was calculated as follows.
We have set up a grid of $400 \times 400$ cells on the 1/4 of the
surface of section $(q,p)$, thus reduced due to the symmetries
(reflection symmetry and time reversal symmetry). The grid points are 
defined  as  $(q_i,p_j) = (\Delta q/2+i\Delta q, \Delta p/2+j\Delta p)$,
where $\Delta q= {\cal L}/800$ and $\Delta p =1/400$. The grid
covers the 1/4 of the entire surface of section, namely
$q\in [0,{\cal L}/2]$ and $p\in [0,1]$. The grid points are positioned at the
centers of the square cells of the area $\Delta q\Delta p$. 
The integration method used to evaluate (\ref{Husfun}) is a 
simple trapeze rule with the step $ds \propto \la_B/20$, where
$\la_B=2\pi/k_j$ is the de Broglie wavelength.
The important point is now that the values of the Poincar\'e Husimi functions
on the grid are normalized in such a way that their sum
is equal to one.

\begin{figure}
\begin{center}
 \includegraphics[width=7.5cm]{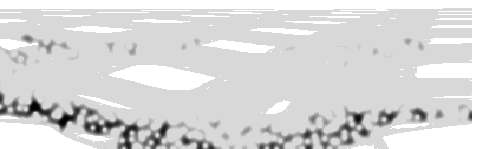}
 \includegraphics[width=7.5cm]{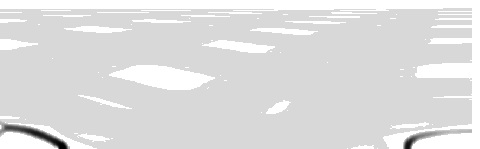}
 \includegraphics[width=7.5cm]{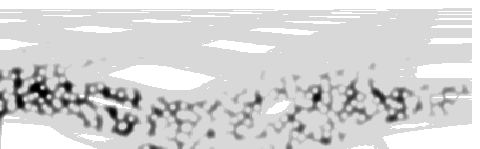}
\includegraphics[width=7.5cm]{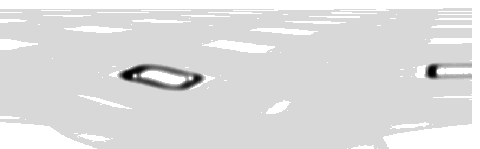}
 \includegraphics[width=7.5cm]{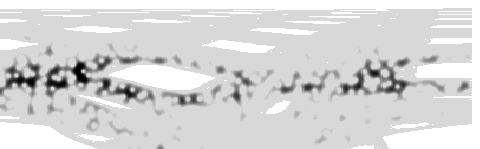}
 \includegraphics[width=7.5cm]{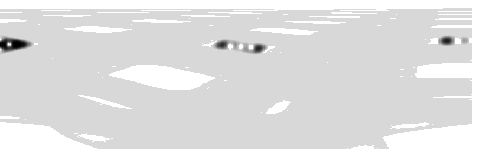}
 \includegraphics[width=7.5cm]{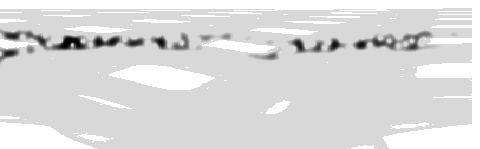}
 \includegraphics[width=7.5cm]{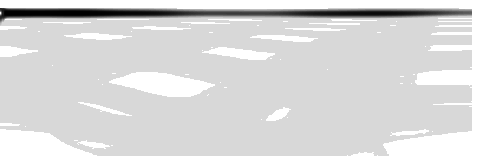}
 \caption{Examples of chaotic (left) and regular (right) states in the Poincar\'e-Husimi
   representation. $k_j$ ($M$) from top down are: chaotic: $k_j$ ($M$) = 2000.0021815 (0.978), 2000.0181794
(0.981), 2000.0000068 (0.989), 2000.0258600 (0.965); 
regular:  $k_j$ ($M$) = 2000.0081402 (-0.987), 2000.0777155 ( -0.821),  2000.0786759 ( -0.528), 2000.0112417 ( -0.829). The gray background 
is the classically chaotic invariant component. 
We show only one quarter of the surface of section  
$(s,p)\in [0,{\cal L}/2] \times [0,1]$, because
due to the reflection symmetry and time-reversal symmetry the
four quadrants are equivalent. } 
\label{fig2}
\end{center}
\end{figure}

Some examples of the Poincar\'e Husimi functions
are shown in figure \ref{fig2}.

Now the classification
of eigenstates can be performed by their projection onto the classical
surface of section. As we are very deep in the semiclassical regime
we do expect with probability one that either an eigenstate is
regular or chaotic, with exceptions having measure zero, ideally.
To automate this task we have ascribed to each point on the grid
a number $A_{i,j}$ whose value is either $+1$ if the grid point lies
within the classical chaotic region or $-1$ if it belongs to a 
classical regular region. Technically, this has been done as follows.
We have taken an initial condition in the chaotic region, and iterated 
it up to about $10^{10}$ collisions, enough for the convergence
(within certain very small distance).
Each visited cell $(i,j)$ on the grid has then been assigned value 
$A_{i,j} = +1$, the remaining ones were assigned the value -1.

The Poincar\'e Husimi function $H(q,p)$
(\ref{Husfun}) (normalized) was calculated on the grid points and the overlap
index $M$ was calculated according to the definition

\be \label{indexM}
M = \sum_{i,j} H_{i,j}\; A_{i,j}.
\ee
In practice, $M$ is not exactly $+1$ or $-1$, but can have a value
in between. The reasons are two, first the finite discretization
of the phase space (the finite size grid), and second,  the
finite wavelength (not sufficiently small effective Planck constant,
for which we can take just $1/k_j$). If so, the question is, where
to cut the distribution of the $M$-values, at the threshold value $M_t$,
such that all states with $M<M_t$ are declared regular and 
those with $M>M_t$ chaotic.

There are two natural criteria: {\bf (I)} {\em The classical criterion:} 
the threshold value $M_t$ is chosen such that we have exactly
$\rho_1$ fraction of regular levels and $\rho_2=1-\rho_1$ of
chaotic levels. {\bf (II)} {\em The quantum criterion:} we choose $M_t$
such that we get the best possible agreement of the chaotic
level spacing distribution with the Brody distribution (\ref{BrodyP}),
which is expected to capture the dynamical localization effects
of the chaotic eigenstates.

\section{Results} \label{Results}

\begin{figure}
\center
\includegraphics[width=7.5cm]{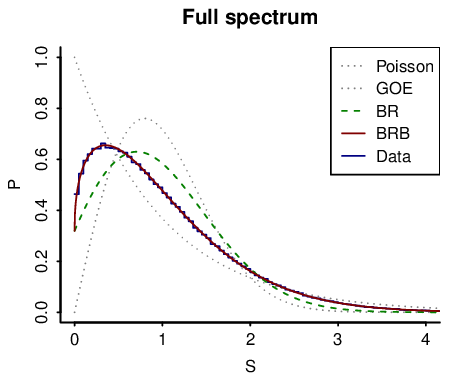}
\includegraphics[width=7.5cm]{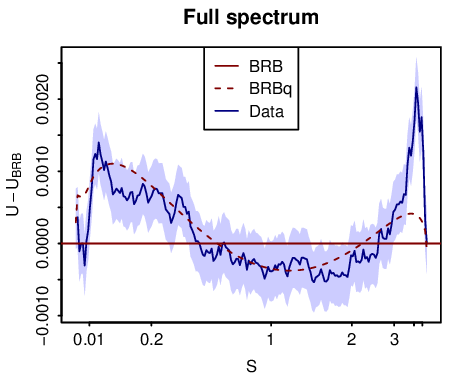}
\caption{(a; left) The level spacing distribution for the 
entire spectrum after unfolding for $N=587653$ spacings,
with $k_j\in [2000,2500]$, 
in excellent agreement with the BRB distribution with
the classical $\rho_1=0.175$ and $\beta=0.45$. 
In the $U$-function plot (b; right), we show $U(data)-U(BRB)$
as a function of $S$, and it is clearly seen that the 
BRB distribution with the quantally determined $\rho_1=0.19$
and $\beta=0.47$ 
is even better fit to the data (dashed, denoted by BRBq).
The belt around the data curve indicates 
the expected statistical $\pm$ one-sigma error.}
\label{fullspectrum}
\end{figure}

To begin with we first look at the total energy spectrum
$E_j=k_j^2$, for $k_j\in [2000,2500]$. 
The spectral unfolding was done using the Weyl 
formula with the perimeter corrections. In figure 
\ref{fullspectrum}a  we show the histogram of $N=587653$
level spacings together with the best fitting
BRB distribution (\ref{BRP}), with $E_r$ being Poissonian
and $E_c$ being the Brody gap probability (\ref{BrodyE}),
derived from (\ref{BrodyP}), with the classical $\rho_1=0.175$,
and the Brody parameter $\beta=0.45$.
For the reference the  BR distribution (\ref{BRexplicit}), 
the Poisson and the GOE level spacing distributions (\ref{PCW})
are shown.
In figure \ref{fullspectrum}b we show the $U$-function
representation of the level spacing distribution, as 
introduced by Prosen and Robnik \cite{ProRob1994b} and defined
in the Appendix B. It was the $U$-function which was used
in finding the best fitting distributions.  
Whilst in the first case the agreement
is perfect, in the  $U$-plot we see that the quantally
adjusted parameter $\rho_1=0.19$ instead of its classical value
$0.175$ leads to even better agreement. In this case $\beta=0.47$.
Please observe that the deviations here are already extremely
small, so the significance of the best fit is of extreme
importance. 

\begin{figure}
\center
\includegraphics[width=7.5cm]{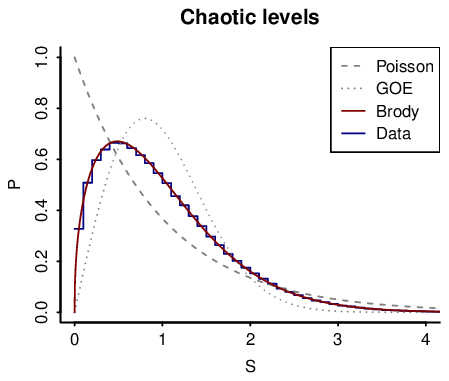}
\includegraphics[width=7.5cm]{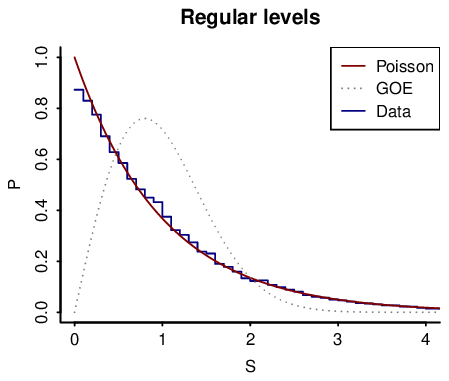}
\caption{Separation of levels using the classical
criterion $M_t=0.431$. (a; left) The level spacing distribution for the 
chaotic subspectrum after unfolding, in perfect agreement with
the Brody distribution $\beta=0.444$.
(b; right) The level spacing distribution for the regular
part of the spectrum, after unfolding, in excellent
agreement with Poisson.}
\label{separspectrum}
\end{figure}

Let us now separate the regular and chaotic eigenstates and
the corresponding eigenvalues, after unfolding, according
to the method described in section \ref{Separ}, using
the classical criterion (I). The
corresponding threshold value of the index $M$ is found
to be $M_t=0.431$. The level spacing distributions are
shown in figure \ref{separspectrum}. As we see, we have perfect
Brody distribution with $\beta=0.444$ for the chaotic
levels and almost pure Poisson for the regular levels.

\begin{figure}
\center
\includegraphics[width=7.5cm]{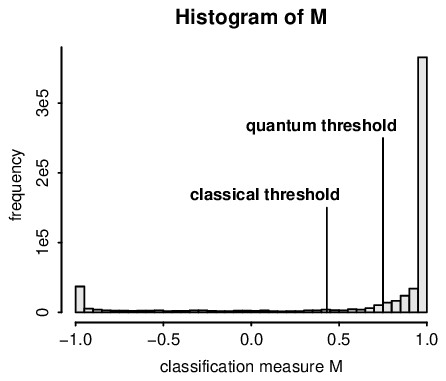}
\includegraphics[width=7.5cm]{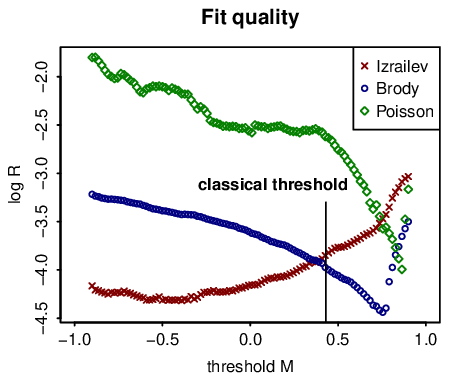}
\caption{(a; left) Distribution of the index $M$ with the locations
of the threshold values of $M$, the 
classical $M_t=0.431$ and the quantum one $M_t=0.75$.
In the first case classical $\rho_1=0.175$, whilst
in the second case the quantal $\rho_1=0.223$.
(b; right) The logarithm of the fit deviation  measure 
$\log_{10} R$, defined in (\ref{devmeas}),
for the Brody and Izrailev distributions for chaotic
levels $M>M_t$,
versus $M_t$, and the same for Poisson distribution for
the regular levels with $M<M_t$. }
\label{histoM}
\end{figure}

In order to make our analysis deeper and more refined
we show first the histogram of the $M$-values in figure
\ref{histoM}a, where we see the two different 
threshold values $M_t$, namely the classical one at $M_t=0.431$,
corresponding to $\rho_1=0.175$,
and the quantum one at $M_t=0.75$ in which case we get
$\rho_1=0.223$.  In figure \ref{histoM}b we analyze the
goodness of the two fits, Brody and Izrailev. First we
define the fit deviation measure as

\be \label{devmeas}
R = \int_0^{\infty} P(S)^2 \left((P(S) - P_F(S)\right)^2\; dS,
\ee
where $P(S)$ describes the data, whilst $P_F(S)$ denotes
the theoretical distribution fitting the data, namely
$F$ stands for Brody or Izrailev. At each chosen threshold value
$M_t$ we consider the set of chaotic levels for which
by definition $M\ge M_t$. By performing the best fit
at such $M_t$ (= threshold $M$) both for Brody and Izrailev,
we calculate the fit deviation measure (\ref{devmeas}) 
and plot its decadic logarithm 
as a function of $M_t$ in figure \ref{histoM}b.
We see that at low $M_t$ Izrailev is somewhat better than 
Brody, but this is the unphysical domain of much too
small $M_t$. Near the classical threshold $M_t=0.431$ they become
comparably good, but at increasing the $M_t$ Brody exhibits
a very sharp and narrow minimum, whilst Izrailev curve
increases. This sharp minimum of $R$ for Brody is at the value 
of $M$ which by definition we called the quantum threshold $M_t=0.75$.
In addition, we show the $R$ quantity (\ref{devmeas}) 
for the Poisson distributioin vs. $M_t$, where we see
also a deep sharp minimum at $M_t\approx 0.8$, thus 
almost at the quantum threshold $M_t=0.75$. 
The conclusion of this analysis is that by varying
the threshold value $M_t$ there is a point $M_t=0.75$
at which the Brody distribution for all chaotic levels with
$M>M_t$ is globally the best  and also better than Izrailev 
fit at any $M_t$. For logical consistency, it is important
that at (almost) the same value of $M_t$ the fit of
the Poisson distribution for the regular levels with $M<M_t$ 
is globally the best, as is evident from the $R$ plot
in figure \ref{histoM}b.

\begin{figure}
\center
\includegraphics[width=7.5cm]{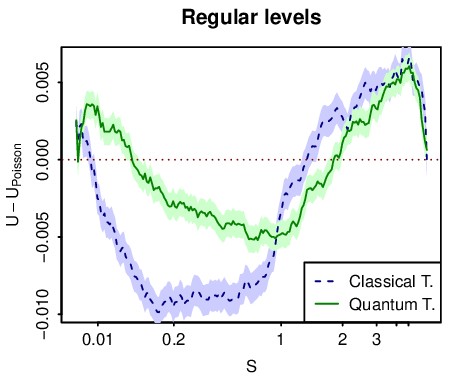}
\includegraphics[width=7.5cm]{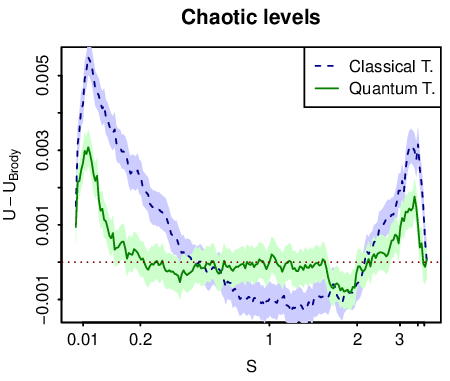}
\caption{The $U$-function plots as differences
$U(data)-U(ideal)$ for the regular and
chaotic levels, for both criteria, the classical one
and the quantum one. The belts around  the data lines
indicate the expected statistical $\pm$ one-sigma errors.}
\label{Uregcha}
\end{figure}

We show the $U$-function plots for the regular levels
and the chaotic levels in figure \ref{Uregcha}, using the
Brody distribution, for
the classical and quantum criterions, corresponding 
to the two different values of $M_t$ explained above.
In both cases we plot the difference $U(data)-U(ideal)$,
so that in case of perfect agreement the line would 
coincide with the abscissa. We clearly see that the
quantum criterion yields noticable better agreement than
the classical criterion. Again, it must be emphasized
that the agreement is extremely good, and the deviations 
of $U(data)$ from $U(ideal)$ are very small numbers.

Finally, we should comment on the relevance of the Brody
distribution. The Brody distribution \cite{Bro1973,Bro1981}
(\ref{BrodyP})
still has no theoretical foundation, but it definitely
captures correctly the effects of dynamical (Chirikov)
localization of chaotic eigenstates. This has been recently
confirmed by Manos and Robnik \cite{MR2012} in case of
the kicked rotator, namely for the quasienergies,
and clearly is demonstrated in the present work for the
autonomous Hamilton system, exemplified by the
2D billiard that we have chosen for this analysis.
It remains as an open theoretical problem to derive the
Brody distribution in this context.

\begin{figure}
\center
\includegraphics[width=7.5cm]{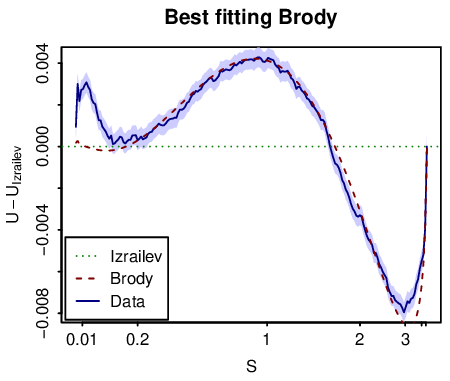}
\includegraphics[width=7.5cm]{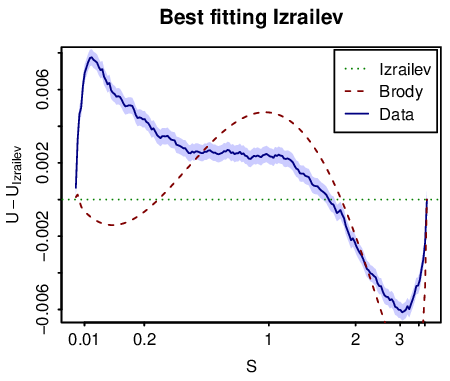}
\caption{We show the $U$-function plot for the
chaotic levels, clearly showing that
Brody distribution (dashed) is much better
than Izrailev distribution. On the left we
show the best fitting Izrailev distribution at the 
point where Brody is globally the best fit using the quantum
threshold $M_t=0.75$. On the right we show 
the Izrailev fit at the point $M_t=-0.5$ where
it is globally the best, and the best Brody fit at the same
$M_t=-0.5$, showing that also there the Brody fit
is better than Izrailev.  This is impressive because the effects
are small, and the statistical significance very high.
The belt around the data line indicates the 
expected statistical $\pm$ one-sigma error.}
\label{Izrailev}
\end{figure}

Since the Brody distribution is not known theoretically
to be the preferred and/or ``the right one", we have considered 
again also the Izrailev distribution \cite{Izr1988,Izr1989,Izr1990} 
(\ref{IzrailevDistrib}), studied
very recently also in \cite{MR2012}. Entirely
in line with the findings in \cite{MR2012} we found that
Brody is much better model of the level spacing distribution
than the Izrailev's one. This is clearly demonstrated in
figure \ref{Izrailev}.

\section{Conclusions} \label{Conclusion}

We have used a billiard system of the mixed type (\ref{defbil}),
with $\la=0.15$, as introduced in \cite{Rob1983}-\cite{Rob1984},
and have shown that using the Poincar\'e Husimi functions we can
separate the regular and chaotic eigenstates. 
The successful separation of course also entirely confirms the
Berry-Robnik picture \cite{BR1984} of separating the regular
and chaotic levels in the semiclassical limit, where
the tunneling effects can be neglected.
With great and  unprecedented statistical
significance we have shown that the chaotic levels exhibit Brody level
spacing distribution, whilst the regular levels obey Poissonian
statistics.  This analysis not only confirms the Berry-Robnik
picture \cite{BR1984} of conceptually separating the regular and
chaotic levels, based on the PUSC and embodied in formula (\ref{BRE}), 
but also demonstrates that the dynamical
localization effects of the chaotic eigenstates are very well
captured by the Brody distribution, in analogy with the same
finding in the Floquet systems, in particular the kicked 
rotator \cite{Izr1990,MR2012}, where the quasienergy spectra
are analyzed. One educated guess for the occurrence of
fractional power law level repulsion (meaning $0 < \beta < 1$)
in dynamically localized but classically fully chaotic periodic
systems is Izrailev's observation (see e.g. \cite{Izr1990} and the
references therein), that the joint probability
distribution for a Circular Orthogonal Ensemble (COE), which
so far as the level spacings are concerned is also GOE, 
should be generalized in the sense of Dyson for noninteger
$\beta$. Of course, this is just a hypothesis, and it
certainly cannot indicate theoretically whether Brody or Izrailev
distribution should be "the right one", leaving us with
the empirical studies and conclusions of this work.
The analogies and connections between the billiard problem
and the Floquet problem have been discussed in more detail
in the recent work \cite{BMR2013}.

Whilst in the kicked rotator the relationship
between the localization measure of the eigenstates and
the spectral level repulsion (Brody) parameter $\beta$ exists,
as proposed by Izrailev, and confirmed by Manos and Robnik,
in the time independent Hamilton systems, like the one discussed
in the present work, such relationship is lacking and is open
for the future work. It involves great numerical efforts.

The theoretical derivation of the Brody
level spacing distribution for the dynamically localized
eigenstates  is thus also an open problem for the future. The billiard
systems are  not just nice theoretical toy models,
but are suitable also for the experimental applications,
like in quantum dots, and microwave cavities introduced and
studied extensively over decades by H.-J. St\"ockmann \cite{Stoe}. 
We also propose to study from the present point of view
the hydrogen atom in strong magnetic field
as an example of classical and quantum chaos par excellence,
as introduced in \cite{Rob1981,Rob1982,HRW1989,WF1989,RWHG1994},
although the technical efforts to obtain large stretches of high-lying
eigenstates and the corresponding energy levels 
are much bigger than in billiard
systems, where we have a great number of different elegant
numerical techniques \cite{VPR2007}, all of them used in our
recent work \cite{BatRob2010}.

\section*{Acknowledgements}

Financial support of the Slovenian Research Agency ARRS
under the grants P1-0306 and J1-4004 is gratefully acknowledged.

\section*{Appendix A: The semiclassical condition}

\begin{figure}
\center
\includegraphics[width=15cm]{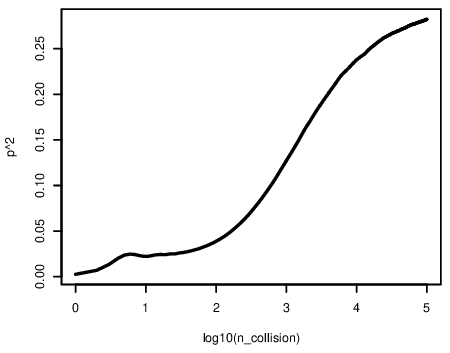}
\caption{We show the second moment $\langle p^2 \rangle$ averaged
over an ensemble of 2000 initial conditions  uniformly distributed
in the chaotic component on the interval $s\in [0,{\cal L}/2]$ and $p=0$ as a function of 
the decadic logarithm of  the 
number of collisions. We see that the saturation value of 
$\langle p^2 \rangle$ is reached at about $N_T=10^5$ collisions. }
\label{trans}
\end{figure}

Here we calculate the Heisenberg time and the classical
transport time for the billiard domain ${\cal B}$ defined in  
equation (\ref{defbil}) with $\lambda =0.15$. According to the
leading order of the Weyl formula, which is in fact just the simple
Thomas-Fermi rule, we have for the number of levels $N(E)$ below
and up to the energy $E$ of a Hamiltonian $H({\bf q},{\bf p})$

\be \label{A1}
N(E) = \frac{1}{(2\pi\hbar)^2} \int_{H({\bf q},{\bf p})\le E} d^2{\bf q}\;d^2{\bf p}.
\ee
Since $H= {\bf p}^2/(2m)$, with constant zero potential energy
inside ${\cal B}$, where $m$ is the mass of the billiard
point particle, and $H$ is infinite on the boundary $\partial {\cal B}$,
we get at once

\be \label{A2}
N(E) = \frac{2\pi {\cal A}mE}{(2\pi\hbar)^2}.
\ee
The density of levels is $\rho (E) = 1/(\Delta E) = 
dN(E)/dE = {\cal A}m/(2\pi\hbar^2)$
and thus the Heisenberg time is

\be  \label{A3}
t_H = 2\pi\hbar \rho(E) = \frac{{\cal A} m}{\hbar}.
\ee
The classical transport time is denoted by $t_T$, and in units of
the number of collisions $N_T$ can be written as 

\be \label{A4}
t_T = \frac{\bar{l} N_T}{v} = \frac{ \bar{l}N_T}{\sqrt{2E/m}},
\ee
where $\bar{l}$ is the mean free path of the billiard particle
and $v =\sqrt{2E/m}$ is its speed at the energy $E$. Thus for 
the ratio  $\alpha = t_H/t_T$ we get

\be \label{A5}
\alpha = \frac{t_H}{t_T} = \frac{{\cal A} k}{N_T \bar{l}} 
\ee
where $k= \sqrt{2mE/\hbar^2}$. Taking into account that
$\bar{l} \approx \pi {\cal A}/{\cal L}$ (this is so-called Santalo's
formula, see e.g. \cite{San1976}), we have

\be \label{A6}
\alpha = \frac{t_H}{t_T} = \frac{{\cal L}k}{\pi N_T}.
\ee
This is a general formula valid for any billiard.
In our case ${\cal L} \approx 2\pi$ and we arrive at the
final estimate

\be \label{A7}
\alpha = \frac{2k}{N_T}.
\ee
Thus the condition for the occurrence of dynamical localization
$\alpha \le 1$ is now expressed in the inequality

\be \label{A8}
k \le \frac{N_T}{2}.
\ee
As our levels are in the interval $k\in [2000,2500]$,
and since $N_T$ is estimated as $N_T \approx 10^5$, we see
that the semiclassical condition (\ref{A8}) is very well satisfied.
In figure \ref{trans} we explicitly show the growth of the second moment
of $p$, namely $\langle p^2 \rangle$, for an ensemble of 2000 initial
conditions uniformly distributed in the chaotic component on 
the interval $s\in [0,{\cal L}/2]$  with $p=0$, where the averaging is
taken over the ensemble and the time. We see that indeed
about $N_T \approx 10^5$ collisions are necessary to reach the
saturation value of $\langle p^2 \rangle$. More detailed study
of the relevant transport properties will be published elsewhere
\cite{BatRob2013}.

\section*{Appendix B: The U-function representation of the level spacing distribution}

First we estimate the expected fluctuation (error) of the
cumulative (integrated) level spacing distribution $W(S)$,
which contains $N_s$ objects.
At a certain $S$ we have the probability $W$ that a level
is in the interval $[0,W]$ and $1-W$ that it is in the
interval $[W,1]$. Assuming binomial probability distribution
$P(k)$ of having $k$ levels in the first and $N_s-k$ levels in the
second interval we have

\be \label{B1}
P(k) = \frac{N_s!}{k! (N_s-k)!} W^k (1-W)^{N_s-k}.
\ee
Then the average values are equal to

\be \label{B2}
<k> = N_sW, \;\;\; <k^2> = N_sW + N_s(N_s-1) W^2,
\ee
and the variance 

\be \label{B3}
V(k) = <k^2> - <k>^2  = N_s W (1-W).
\ee
But the probability $W$ is estimated in the mean as $k/N_s$.
Its variance is

\be \label{B4}
V(W) = V\left( \frac{k}{N_s} \right) = \frac{1}{N_s^2} V(k) =  \frac{W(1-W)}{N_s}
\ee
and therefore the estimated error of $W$ (standard deviation,
the square root of the variance) is given by

\be \label{B5}
\delta W = \sqrt {V(W)} = \sqrt {\frac{W(1-W)}{N_s} }.
\ee
Transforming now from $W(S)$ to

\be \label{B6}
U(S) = \frac{2}{\pi} \arccos \sqrt{1 - W(S)},
\ee
we show in a straightforward manner that

\be \label{B7}
\delta U = \frac{1}{\pi \sqrt{ N_s}}
\ee
and is indeed independent of $S$.
From the (choice of the constant pre-factor in the) definition (\ref{B6})
one sees that both $U(S)$ and $W(S)$ go from $0$ to $1$
as $S$ goes from $0$ to infinity.

\end{document}